# Jupiter – friend or foe? II: the Centaurs

**J. Horner and B. W. Jones**

*Astronomy Group, Physics & Astronomy, The Open University, Milton Keynes, MK7 6AA, UK*

*e-mail: j.a.horner@open.ac.uk   Phone: +44 1908 653229        Fax: +44 1908 654192*

(SHORT TITLE: Jupiter – friend or foe? II: the Centaurs)

**Abstract**

It has long been assumed that the planet Jupiter acts as a giant shield, significantly lowering the impact rate of minor bodies upon the Earth, and thus enabling the development and evolution of life in a collisional environment which is not overly hostile. In other words, it is thought that thanks to Jupiter, mass extinctions have been sufficiently infrequent that the biosphere has been able to diversify and prosper. However, in the past, little work has been carried out to examine the validity of this idea. In the second of a series of papers, we examine the degree to which the impact risk resulting from objects on Centaur-like orbits is affected by the presence of a giant planet, in an attempt to fully understand the impact regime under which life on Earth has developed. The Centaurs are a population of ice-rich bodies which move on dynamically unstable orbits in the outer Solar system. The largest Centaurs known are several hundred kilometres in diameter, and it is certain that a great number of kilometre or sub-kilometre sized Centaurs still await discovery. These objects move on orbits which bring them closer to the Sun than Neptune, although they remain beyond the orbit of Jupiter at all times, and have their origins in the vast reservoir of debris known as the Edgeworth-Kuiper belt that extends beyond Neptune. Over time, the giant planets perturb the Centaurs, sending a significant fraction into the inner Solar System where they become visible as short-period comets. In this work, we obtain results which show that the presence of a giant planet can act to significantly change the impact rate of short-period comets on the Earth, and that such planets often actually increase the impact flux greatly over that which would be expected were a giant planet not present.

**Key words:** Solar System – general, comets – general, Centaurs, minor planets, planets and satellites – general, Solar System – formation.

**Introduction**

In our previous paper, "Jupiter – friend or foe? I: the asteroids" (Horner & Jones, 2008, Paper I), we highlighted the idea that Jupiter has significantly reduced the impact rate on the Earth of minor bodies, notably small asteroids and comets, thereby allowing the biosphere to survive and develop (for example, see Greaves, 2006). This idea is widely accepted, both in the scientific community and beyond. It is clearly the case that a sufficiently high rate of large impacts would result in the evolution of a biosphere being stunted by frequent mass extinctions, each bordering on global sterilisation. Were Jupiter not present in our Solar System, it is argued, such frequent mass extinctions would occur on the Earth, and therefore the development of life would be prevented.

We also pointed out that, until recently, very little work had been carried out to examine the effects of giant planets on the flux of minor bodies through the inner Solar System. Wetherill (1994) showed that in systems containing bodies which grew only to the size of, say, Uranus and Neptune, the impact flux from comets originating in the Oort Cloud[1], experienced by any terrestrial planet, would be a factor of a thousand times greater than that seen today in our System, as a direct result of less efficient ejection of material from the System during its early days. This work is discussed in more detail in Paper I, which also outlines recent work by Laasko et al. (2006), who conclude that Jupiter "*in its current orbit, may provide a minimal of protection to the Earth*". Paper I also mentions the work of Gomes et al. (2005), from which it is clear that removing Jupiter from our Solar System would result in far fewer impacts on the Earth by lessening, or removing entirely, the effects of the Late Heavy Bombardment in the inner Solar System.

Thus, it seems that the idea of "Jupiter, the protector" dates back to the time when the main impact risk to the Earth was thought to arise from the Oort cloud comets (Wetherill, 1994). Many such objects are actually expelled from the Solar System after their first pass through its inner reaches, as a result of Jovian perturbations, which clearly lowers the chance of one of these cosmic bullets striking the Earth (see, for example, Matese & Lissauer, 2004). Recently, however, it has become accepted that near-Earth objects (primarily asteroids, with a contribution from the short-period comets[2]) pose a far greater threat to the Earth. Indeed, it has been suggested that the total cometary contribution to the impact hazard may be no higher than 25% (Chapman & Morrison, 1994).

In order to study the relationship between a giant planet and the impact rate on a terrestrial world, we are running *n*-body simulations to see how varying the mass of Jupiter would change the impact rate on Earth. Since there are three source populations which provide the main impact threat, the asteroids, the short-period comets, and the Oort cloud comets, we are examining each population in turn. In Paper I we examined the effect of changing Jupiter's mass on the impact rate experienced by the Earth from objects flung inwards from the asteroid belt. Our results were surprising. At very low and very high Jupiter masses, the impact rate was particularly low. However, there was a sharp peak in the impact flux at around 0.20 times the mass of our Jupiter, at which point the Earth in our simulations experienced almost twice as many impacts as it did in the simulation of our own Solar

---

[1] The Oort cloud is a vast shell of icy bodies, centred on the Sun, extending to approximately halfway to the nearest star (some $10^5$ AU). Bodies swung inwards from this cloud typically have orbital periods of tens of thousands, or even millions of years, and are often described as "long-period comets".

[2] Short-period comets typically have orbital periods significantly less than 200 years. In contrast to the long-period comets, the great majority of these objects originate from the Edgeworth-Kuiper Belt (which stretches out to around 20 AU beyond the orbit of Neptune), and from the associated Scattered Disk.

System. This shows conclusively that the idea of "Jupiter – the shield" is far from a complete description of how giant planets affect terrestrial impact fluxes, and that more work is needed to examine the problem.

In this paper, we detail our results for the short-period comets. The main source of these objects is the Centaurs, a transient population of ice-rich bodies ranging up to a few hundred kilometres across. They orbit with perihelia between the orbits of Jupiter and Neptune, and are themselves sourced from the region just beyond the orbit of Neptune, where the Edgeworth-Kuiper belt and the Scattered Disk objects lie (Horner, Evans & Bailey 2004; Levison & Duncan, 1997). The giant planets perturb the Centaurs, and send a significant fraction into the inner Solar System, where they become visible as short-period comets. Our results for Oort cloud comets (the reservoir studied by Wetherill) will be detailed in later work.

**Simulations**

Of the three parent populations for Earth-impacting bodies, the simplest to model are the short-period comets. However, given that we wished to look at the effects of Jupiter on the impact flux, taking a population which has already been significantly perturbed by the giant planet would clearly have been a mistake. Instead, we chose to use the Centaurs to provide our population of potentially threatening objects.

In order to create a swarm of test objects which might evolve onto Earth-impacting orbits, we searched the Centaur and Trans-Neptunian ("Beyond Neptune") object lists hosted by the Minor Planet Center for all objects with perihelia between 17 and 30 AU (see, for example http://www.cfa.harvard.edu/iau/lists/Centaurs.html, http://www.cfa.harvard.edu/iau/lists/TNOs.html). This gave a total of 105 objects, including Pluto. Pluto was removed, giving a sample of 104 objects. These were then "cloned" 1029 times each, with each orbit obtained from the MPC acting as the central point in a 7x7x7x3 grid in *a-e-i-ω* space (with clones separated by 0.1 AU in semi-major axis, 0.05 in eccentricity, 0.5 degrees in inclination, and 5 degrees in the argument of perihelion). The steps used, and the number of clones created in a given element, were chosen to disperse the clones widely enough in orbital element space around the "parent" that rapid dynamical dispersion would occur. In addition, it is clear that our initial sample of 104 objects contains a number of bodies on stable orbits (in mean-motion resonances, for example). Given that we are interested in the behaviour of those objects in the outer Solar system which have already left the stable reservoirs, it was important that the cloning process move many of the clones of these objects onto less stable orbits, allowing them to diffuse through the Solar System within the period of our integrations.

The cloning process produced a population of just over 107000 objects covering a wide range of values in orbital element space, orbits which were simulated for a period of 10 million years using the hybrid integrator contained within the *MERCURY* (Chambers, 1999) package, with an integration time step of 120 days, along with the planets Earth, Jupiter, Saturn, Uranus and Neptune, all with initial orbital elements equal to their present values (though they barely changed during the simulation). The integration length was chosen to provide a balance between required computation time and the statistical significance of the results obtained. In the simulation the cloned objects were treated as massless particles, feeling the gravitational pull of the planets and the Sun, but experiencing no interaction with one another. The massive bodies (the planets), in turn, experienced no perturbation from the massless particles, but were able to fully interact with one another.

As in our Paper I, the Earth within our simulations was inflated to have a radius of one million kilometres, in order to enhance the impact rate from objects on Earth crossing orbits. Simple initial integrations were again carried out to confirm that this inflation did affect the impact rate as expected, with the flux scaling as expected with the cross-sectional area of the planet. In order to examine the effect of Jovian mass on the impact rate, we ran thirteen separate scenarios. In the first, we used a Jupiter with the same mass as that in our Solar System (so one Jupiter mass), while in the others, planets of mass 0.01, 0.05, 0.10, 0.15, 0.20, 0.25, 0.30, 0.50, 0.75, 1.50 and 2.00 times the mass of the present Jupiter were substituted in its place. Finally, a run was carried out in which no Jupiter was present. Hereafter, we refer to these runs by the mass of the planet used, so that, for example, $M_{1.00}$ refers to the run using a planet of one Jupiter mass, and $M_{0.01}$ refers to the run using a planet of 0.01 Jupiter masses. The (initial) orbital elements of "Jupiter", together with all the other planets, were identical in all cases.

Though, in reality, if our Solar System had formed with a Jupiter of different mass, the architecture of the outer Solar System would probably be somewhat different, rather than try to quantify the uncertain effects of a change to the configuration of our own Solar System, we felt it best to change solely the mass of the "Jupiter" in our work, and therefore work with a known, albeit modified, system, rather than a theoretical construct. For a flux of objects moving inwards from the Edgeworth-Kuiper belt, this does not seem unreasonable – by choosing a population of objects well beyond the "Jupiter" in our simulations, with initial perihelia between 17 and 30 AU, we have greatly reduced the planet's influence on the objects prior to the start of our simulations, and believe this method allows us to make a fair assessment of the role of Jovian mass on such objects.

The complete suite of integrations ran for some nine months of real time, spread over the cluster of machines sited at the Open University. This nine months of real time equates to over twelve years of computation time, and resulted in measures of the impact flux for each of the thirteen "Jupiters". Further, the eventual fate of each object was followed, allowing the determination of the dynamical half-life of the population in the different runs. With the constant trickle of objects being lost by ejection or collision with the Sun or with planets other than the Earth, this half-life is clearly an important factor in determining the threat posed, since a more stable population (one with a longer half-life), with the same parent-flux, would lead to an enhanced population of impactors, potentially negating any shielding effect resulting from the lowered impact rate per simulated object.

Note that objects placed on Earth-crossing orbits will de-volatilise on a time scale orders of magnitude shorter than the 10 Myr of our integrations. Comets are observed to fragment or disintegrate during their lifetimes with some regularity (a famous example of such disintegration being comet 3D/Biela, which is discussed in some length in Babadzhanov et al., 1991). However, it seems likely that many comets simply age and "switch off", becoming husks which resemble asteroidal bodies (Levison et al., 2006). It is unlikely, then, that the effects of de-volatilisation will alter the main thrust of our results, even though objects travelling inward from orbits with initial perihelia beyond about 17 AU can remain in the inner Solar System for periods significantly longer than the theoretical de-volatilisation time (Horner & Evans, 2004).

**Results**

As can be seen from Figure 1, the rate at which objects hit the Earth clearly ranges widely as a function of the mass of the Jupiter-like planet in each simulation. The run without a Jupiter ($M_{0.00}$), shown in red on the upper left hand panel, clearly displays a much slower start to the impacts than the runs involving higher mass planets (upper right hand plot). This is a result of the fact that, with no Jupiter to nudge things our way, the bulk of the work is done by Saturn, which, being both smaller and further away from the Earth, has a much harder time injecting Earth-crossers.

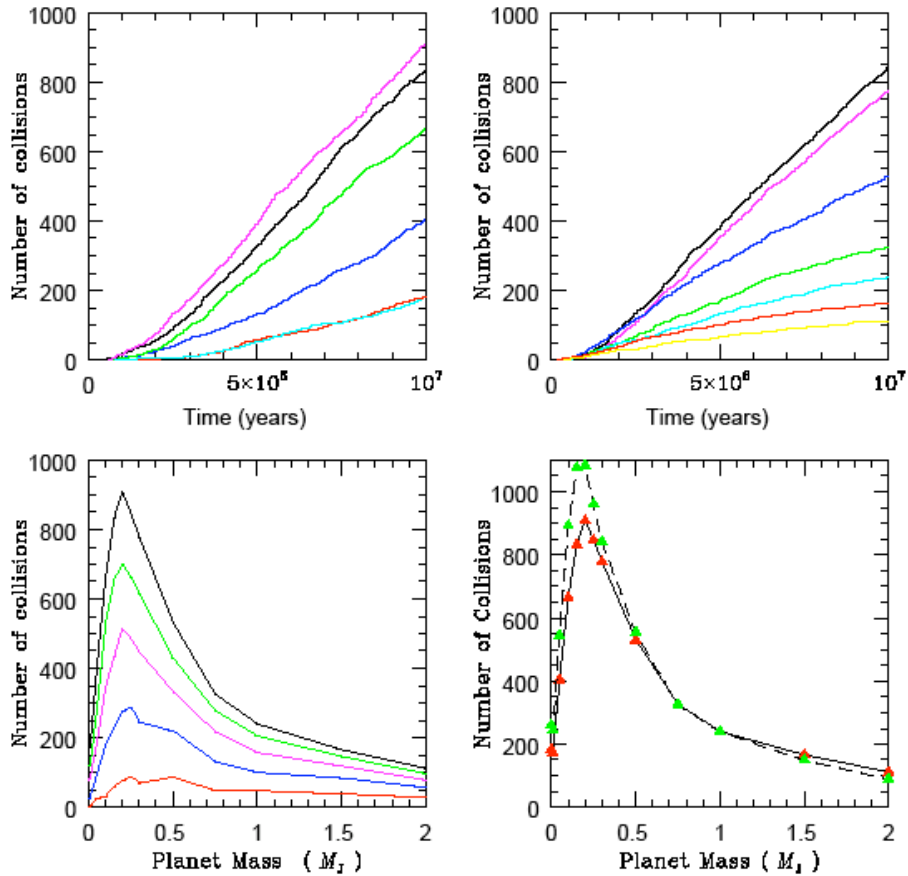

Figure 1   The top two panels show the number of collisions of the simulated objects with the Earth, as a function of time. At left, we have $M_{0.00}$ (red), $M_{0.01}$ (cyan), $M_{0.05}$ (blue), $M_{0.10}$ (green), $M_{0.15}$ (black), and $M_{0.20}$ (magenta). At right, we have $M_{0.25}$ (black), $M_{0.30}$ (magenta), $M_{0.50}$ (blue), $M_{0.75}$ (green), $M_{1.00}$ (cyan), $M_{1.50}$ (red) and $M_{2.00}$ (yellow). The lower left-hand panel gives the number of impacts as a function of the mass of the Jupiter in the simulation: $t$ = 2Myr (red), 4Myr (blue), 6Myr (magenta), 8Myr (green), 10Myr (black). Finally, the panel at the lower right shows the number of impacts as a function of the Jupiter mass after ten million years. The solid line and red triangles show the results from our simulations, while the green triangles and dashed line show these numbers adjusted to take account of the variations in the half-life of the ensemble compared to that of the $M_{1.00}$ simulation.

The third column in Table 1 gives $N_{ejected}$ – the total number of objects that were removed during the course of the simulations. In our runs, objects were destroyed either on impact with one of the massive bodies (the Sun, Earth, Jupiter, Saturn, Uranus and Neptune), or on reaching a distance of 1000 AU from the central body. Unlike the figure showing the impact rate on Earth, it is clear from the tabulated data that the rate at which the objects are removed from the Solar System increases with the mass of Jupiter.

The value of $N_{ejected}$ has been adjusted to take account of the fact that, in each of the runs, 883 of the initial population of objects were placed on orbits so eccentric that they reached the 1000 AU ejection distance on their first orbit. These have been removed from the total in each case, and the value of the dynamical half-life, $T_{1/2}$, has been calculated from this modified value – $T_{1/2}$ is obtained for the corrected population of objects in each simulation, in Myr. The total number of particles used in the calculation has, similarly, been modified, so that $T_{1/2}$ represents that of the 99.9% of our population which started the simulation on bound orbits. One thing that is immediately obvious is that the number of particles removed from the simulations varies far less than the impact flux at Earth, and increases with increasing Jupiter mass. This illustrates the increasing efficiency with which Jupiter flings objects from the Solar System as its mass increases.

Given that systems which display longer dynamical half-lives would be expected to have a larger steady state population, the number of impacts, $N_{impact}$ has been scaled by the ratio of $T_{1/2}$ of the run in question to that in the $M_{1.00}$ case, to give $N_{imp-s}$. This illustrates the effect that Jupiter has in diminishing the particle population by accelerating their ejection from the system, and gives a more realistic view of the changes in impact flux as a function of Jovian mass. The results of this calculation are given in the final column in Table 1, and plotted in the lower right-hand panel of Figure 1.

**Discussion**

From the results discussed above, it is clear that the notion that any "Jupiter" would provide more shielding than none at all is incorrect, at least for impactors originating from the population of small objects with initial perihelia in the range 17-30 AU. It seems that the effect of such a planet on the impact flux on potentially habitable worlds is far more complex than was initially thought. With our current Jupiter, potentially impacting objects are ejected from the system with such rapidity that they pose rather little risk for planets in the habitable zone, and therefore, Jupiter offers a large

Centaurs – IJA  BWJ + JAH  Version for Astro-ph        8 of 13                    19/3/09  9:13 AM UT

degree of shielding, compared to smaller versions of the planet. Planets much more massive than Jupiter clearly offer an even higher degree (as can be seen in the $M_{1.50}$ and $M_{2.00}$ cases).

| $M$ (in $M_J$) | $N_{impact}$ | $N_{ejected}$ | $T_{1/2}$ (in Myr) | $N_{imp-s}$ |
|---|---|---|---|---|
| 0.00 | 181 | 8103 | 87.3 | 259 |
| 0.01 | 172 | 8191 | 86.3 | 243 |
| 0.05 | 403 | 8569 | 82.3 | 543 |
| 0.10 | 664 | 8595 | 82.1 | 892 |
| 0.15 | 832 | 8914 | 79.0 | 1076 |
| 0.20 | 907 | 9612 | 73.0 | 1083 |
| 0.25 | 846 | 10088 | 69.4 | 960 |
| 0.30 | 777 | 10566 | 66.1 | 841 |
| 0.50 | 530 | 10896 | 64.0 | 555 |
| 0.75 | 325 | 11353 | 61.3 | 326 |
| 1.00 | 239 | 11375 | 61.1 | 239 |
| 1.50 | 165 | 13244 | 55.2 | 149 |
| 2.00 | 112 | 14941 | 48.4 | 89 |

Table 1    $M$ is the mass of the "Jupiter" used in a given run (relative to that of the real Jupiter), $N_{impact}$ is the number of impacts on the Earth over the course of the simulation, while $N_{ejected}$ gives the total number of objects removed from the simulation through either collision (with massive bodies other than the Earth) or ejection. The value of $N_{impact}$ is a corrected value, and ignores the 883 objects that were placed on orbits which led to immediate ejection in the initial population process. $T_{1/2}$ gives the dynamical half-life obtained for the corrected population of massless bodies in each simulation, in Myr. $N_{imp-s}$ gives the number of impacts that would be expected from a population enhanced over that described here, as a result of the longer lifetime of the objects in a given system. It is scaled so that the result for the case of our Solar System ($M_{1.00}$) remains the same.

At the other end of the scale, when the "Jupiter" is of particularly small mass (or when no "Jupiter" is present), fewer objects are scattered onto orbits which cross the habitable zone, and so, once again, the impact rate is low. The more interesting situation occurs for intermediate masses, where the giant is massive enough to emplace objects on threatening orbits, but small enough that ejection events are still infrequent. The situation which offers the greatest enhancement to the impact rate is one located around 0.20 $M_J$, in our simulations, at which point the planet is massive enough to efficiently inject objects to Earth-crossing orbits, but small enough that the time spent on these orbits is such that the impact rate is significantly enhanced.

The effect of Jupiter on the size of the population of incoming bodies is one that must be considered together with its direct effect on the impact rate from a population of a given size. Given that our various scenarios feature a Solar System like our own, with only the mass of Jupiter changing, it is clear that the inward flux from beyond Neptune would be unchanged between the various scenarios. However, as the mass of Jupiter falls, the effciency with which the objects are ejected falls, and therefore $T_{1/2}$ of that transient population rises. With a longer $T_{1/2}$, and the same source flux, the total population at a given time would be larger than for cases with shorter $T_{1/2}$, as a steady-state would be reached with more objects moving around the outer Solar System. $T_{1/2}$ of our population in the $M_{0.00}$ case is 87 million years, compared with a value of 61 Myr in the $M_{1.00}$ case, which is readily shown to mean that the true population in the $M_{0.00}$ case would be some 40% higher than that in the $M_{1.00}$ case. Given that the impact rate should scale linearly with the population of objects, it is clear that this means that the impact rate in the $M_{0.00}$ case should be scaled upwards by some 40% to be directly comparable to the $M_{1.00}$ case. The results of such calculations are shown in both Figure 1 and Table 1. In the case of our $M_{0.00}$ integrations, 180 simulated objects hit the Earth, compared to 240 collisions in the $M_{1.00}$ simulation. If we modify the impact rates as described above, the situation is changed from 180:240 to 260:240.

From this we can see that, once one takes into account the increased stability of the object population in the low-mass runs, our Jupiter is actually almost equivalently as effective a shield as having none at all, rather than appearing slightly more threatening than the Jupiter-free case. For the higher masses, the half-lives are close enough that the difference in population will be fairly minor, but it is important to keep this population enhancement in mind for the runs with lower "Jupiter" masses.

As the mass of "Jupiter" decreases towards that of Saturn, 0.30 $M_J$, the relative influence of the planet Saturn in converting Centaurs into short-period comets clearly becomes ever greater. That said, since Saturn is almost twice as far from the Sun as Jupiter, and therefore finds it significantly more difficult to implant objects to Earth-crossing orbits, the simulated Jupiter continues to be the dominant source of such comets down to particularly low masses. At yet smaller masses, approaching that of the planet Mars, 0.00034 $M_J$, one might ask whether we were justified in ignoring the effect of the planet Mars in our simulations. We contend that Mars plays little or no role in the delivery of the great majority of cometary material to the inner Solar System, even in cases where Jupiter is not present. Without Jupiter, Saturn plays the key role in the injection of such bodies – Mars is so small that its effect on passing bodies is negligible, and objects placed on Mars-crossing orbits by Jupiter or Saturn are generally far more likely to be moved onto Earth-crossing orbits by one of the giant planets than by their minute sibling.

**Conclusions**

As pointed out in Paper I, the idea that the planet Jupiter has acted as an impact shield through the Earth's history is one that is entrenched in planetary science, even though little work has been done to examine this idea. In the second of an ongoing series of studies, we have examined the question of Jovian shielding using a test population of particles on orbits representative of the Centaurs and those trans-Neptunian objects with perihelia between 17 and 30 AU, icy bodies that represent the parent population of the short-period comets (through pathways similar to those described in Horner et al., 2003). This is one of three reservoirs of potentially hazardous objects (our Paper I deals with objects sourced from the asteroid belt, and a future paper will deal comets swung inward from the Oort cloud.)

For the studied population, it seems that the Solar System containing our Jupiter is only about as effective in shielding the Earth as a system containing no Jupiter at all. Furthermore, it seems that terrestrial planets in systems containing smaller "Jupiters" would be subject to a significantly higher rate of impacts than those in systems with planets larger than our own Jupiter. Though our work currently studies systems which differ from our own only in the mass of "Jupiter", the broad terms in which it is stated could well apply widely. Only further work will tell, but our initial results already offer intriguing hints as to the true role of giants in the determination of planetary habitability.

This work is doubly interesting when considered in concert with the results we obtained in our previous paper (Horner & Jones, 2008 – Paper I), which reports the effect of Jupiter on impactors from the asteroid belt. As stated in the Introduction, we found that "Jupiters" of low and high mass caused fewer impacts than those of intermediate mass ($M \sim 0.2\ M_J$), with a similar sharp rise and fall from the impact-maximum to that observed in this paper. In both works, we find that planets of mass similar to, or a bit smaller than, the planet Saturn pose the greatest threat to terrestrial worlds in planetary systems like our own, when placed at Jupiter's current location. For the asteroids, we concluded that this was primarily a result of the depth, breadth and location of the $\nu_6$ secular resonance in the main asteroid belt, while for the short-period comets it seems to be down to the interplay between the injection rate of Earth-crossers with the efficiency with which they are then removed from the system. Despite the different causes, the similarity between the shapes of the impact distributions is striking.

Future work will continue the study of the role of Jupiter in limiting or enhancing the impact rate on the Earth by examining bodies representative of the Oort cloud (source of the long-period comets, the population of potential impactors studied by Wetherill in 1994), together with examining the effect of Jovian location on the impact fluxes engendered by the three populations. Given the surprising outcome of our work to date, we hesitate to anticipate future outcomes.

Additionally, future work will also consider whether the absence of a Jupiter-like body would change the populations of objects which reside in the reservoirs that provide the bulk of the impact hazard, a possible effect ignored in this work.

**Acknowledgements**

This work was carried out with funding from STFC, and JH and BWJ gratefully acknowledge the financial support given by that body.